\documentclass[preprint,11pt,amsmath,amssymb]{revtex4}
\usepackage{epsfig}
\usepackage{graphicx}

\begin{document}

\title{Primordial power spectrum of tensor perturbations in Finsler spacetime}

\author{Xin Li $^{1,2}$}
\email{lixin1981@cqu.edu.cn}
\author{Sai Wang $^2$}
\email{wangsai@itp.ac.cn}
\affiliation{$^1$Department of Physics, Chongqing University, Chongqing 401331, China\\
$^2$State Key Laboratory Theoretical Physics, Institute of Theoretical Physics, Chinese
Academy of Sciences, Beijing 100190, China}

\begin{abstract}
We first investigate the gravitational wave in the flat Finsler spacetime. In the Finslerian universe, we derive the perturbed gravitational field equation with tensor perturbations. The Finslerian background spacetime breaks rotational symmetry and induces parity violation. Then we obtain the modified primordial power spectrum of tensor perturbations. The parity violation feature requires that the anisotropic effect contributes to $TT,TE,EE,BB$ angular correlation coefficients with $l'=l+1$ and $TB,EB$ with $l'=l$. The numerical results show that the anisotropic contributions to angular correlation coefficients depend on $m$, and $TE$ and $ET$ angular correlation coefficients are different.
\end{abstract}

\maketitle
\section{Introduction}
Symmetry plays an essential role in studying cosmological physics. Cosmic inflation \cite{Starobinsky}, as one of basic ideas of modern cosmology, can be described by nearly de Sitter (dS) spacetime. The nearly dS spacetime preserves the symmetry of spatial rotations and translations. The primordial power spectrum is scale-invariant if the symmetry of time translation of dS spacetime is preserves. The recent astronomical observations on the anisotropy of cosmic microwave background (CMB) \cite{CMB2} show that the exact scale invariance of the scalar perturbation is broken with more than 5 standard deviations. The observations \cite{CMB2} give stringent limit on the magnitude of deviation from the scale invariant, i.e., $\mathcal{O}(10^{-2})$. It means that the primordial power spectrum for scalar perturbation is approximately scale invariant and the symmetry of time translation is slightly broken.

Recently, the CMB power asymmetry has been reported \cite{Power asymmetry}. One possible physical mechanism that accounts for CMB power asymmetry is anisotropic inflation models where the rotational symmetry of the nearly dS spacetime is violated. To induce the anisotropy in inflation, the popular approach is to involve a vector field \cite{Vector model} that aligned in a preferred direction. In such anisotropic inflation model, the comoving curvature perturbation becomes statistically anisotropic \cite{Vector model1}. Usually, the background spacetime of the anisotropic inflation model is described by Bianchi spacetime \cite{Bianchi spacetime}.

Instead of choosing the Bianchi spacetime as a backgroud spacetime, we will use Finsler spacetime \cite{Book by Bao} as a background spacetime to study anisotropic inflation. In general, Finsler spacetime admits less Killing vectors than Riemann spacetime does \cite{Finsler PF}. Also, there are types of Finsler spacetime that are non-reversible under parity flip, $x\rightarrow-x$. A typical non-reversible Finsler spacetime is Randers spacetime \cite{Randers}. Such a property makes a function in Fourier space $\phi(-\vec{k})$ to be different from $\phi^\ast(\vec{k})$. Therefore, in Finsler spacetime, the spatial rotational symmetry and parity symmetry are violated. In Ref.\cite{Finsler scalar modes} , we proposed an anisotropic inflation model in Finsler spacetime. We studied the primordial scalar perturbations, and obtained off-diagonal angular correlations for the CMB temperature fluctuation and E-mode polarization.

In this paper, we apply the Finslerian background spacetime that used in Ref.\cite{Finsler scalar modes} to study the possible modulation in the amplitude of tensor perturbations. In standard model, the $TB$ and $EB$ correlations vanish. This is due to the fact that the parity of the CMB temperature fluctuation and E-mode polarization are different from that of B-mode polarization. However, this is not the case in Finslerian anisotropic inflation model. In Ref.~\cite{Finsler scalar modes}, we have shown that the parity violation feature requires that the anisotropic effect of the primordial power spectrum of scalar perturbations appears in angular correlation coefficients with $l'=l+1$. It means that the anisotropic part of the temperature fluctuations has the same parity with the B-mode polarization. Thus, one can expect that the angular correlations $TB$ and $EB$ have non--vanished value.

This paper is organised as follows. In Section \ref{sec:flat GW}, we investigate the gravitational wave in flat Finsler spacetime. The plane--wave solution of gravitational wave is given by imposing three constraints. In Section \ref{sec:FRW GW}, we investigate tensor perturbations for the modified Friedmann-Robertson-Walker (FRW) spacetime in which the spatial part is replaced by Randers space. In the modified FRW spacetime, we derive the gravitational field equation for the gravitational wave and obtain the primordial power spectrum for gravitational wave. In Section \ref{sec:numerical results}, the angular correlation coefficients for tensor perturbations are given. And we plot the numerical results of the angular correlation coefficients that describes the anisotropic effect. Conclusions and remarks are given in Section \ref{sec:conclusion}.

\section{Gravitational wave in flat Finsler spacetime}\label{sec:flat GW}
Finsler geometry is based on the so called Finsler structure $F$ defined on the tangent bundle of a manifold $M$, with the property $F(x,\lambda y)=\lambda F(x,y)$ for all $\lambda>0$, where $x\in M$ represents position and $y$ represents velocity. The Finslerian metric is given as \cite{Book by Bao}
\begin{equation}
g_{\mu\nu}\equiv\frac{\partial}{\partial
y^\mu}\frac{\partial}{\partial y^\nu}\left(\frac{1}{2}F^2\right).
\end{equation}
The Finslerian metric reduces to Riemannian metric, if $F^2$ is quadratic in $y$. A Finslerian metric is said to be locally Minkowskian if at every point, there is a local coordinate system, such that $F=F(y)$ is independent of the position $x$ \cite{Book by Bao}. It can be proved that all types of curvature tensors vanish in locally Minkowskian spacetime. Thus, the locally Minkowskian spacetime is flat Finsler spacetime.
Throughout this paper, the indices are lowered and raised by $g_{\mu\nu}$ and its inverse matrix $g^{\mu\nu}$.

In Finsler geometry, there is a geometrical invariant quantity, i.e., Ricci scalar. It is of the form \cite{Book by Bao}
\begin{equation}\label{Ricci scalar}
Ric\equiv\frac{1}{F^2}\left(2\frac{\partial G^\mu}{\partial x^\mu}-y^\lambda\frac{\partial^2 G^\mu}{\partial x^\lambda\partial y^\mu}+2G^\lambda\frac{\partial^2 G^\mu}{\partial y^\lambda\partial y^\mu}-\frac{\partial G^\mu}{\partial y^\lambda}\frac{\partial G^\lambda}{\partial y^\mu}\right),
\end{equation}
where $G^\mu$ is geodesic spray coefficients
\begin{equation}
\label{geodesic spray}
G^\mu=\frac{1}{4}g^{\mu\nu}\left(\frac{\partial^2 F^2}{\partial x^\lambda \partial y^\nu}y^\lambda-\frac{\partial F^2}{\partial x^\nu}\right).
\end{equation}
The Ricci scalar only depends on the Finsler structure $F$ and is insensitive to connections.

There are types of gravitational field equation in Finsler spacetime \cite{Li Berwald,Miron,Rutz1,Vacaru,Vacaru ref,Pfeifer}. These gravitational field equations are not equivalent to each other. It is well known that there is only a torsion free connection, i.e., the Christoffel connection in Riemann geometry. However, there are types of connection in Finsler geometry. Therefore, the gravitational field equations that depend on the connection should not be equal to each other. Thus, one should construct the gravitational field equation from geometrical invariant quantity in Finsler spacetime. The analogy between geodesic deviation equations in Finsler spacetime and Riemann spacetime gives the vacuum field equation in Finsler gravity \cite{Finsler Bullet,Finsler BH}. It is the vanishing of Ricci scalar. The vanishing of the Ricci scalar implies that the geodesic rays are parallel to each other. The geometric invariant property of Ricci scalar implies that the vacuum field equation is insensitive to the connection, which is an essential physical requirement.

Before studying the primordial tensor modes in Finslerian inflation model, we investigate the property of gravitational wave in flat Finsler spacetime. We suppose the Finslerian metric is close to the locally Minkowski metric $\eta_{\mu\nu}(y)$,
\begin{equation}\label{flat GW}
g_{\mu\nu}=\eta_{\mu\nu}(y)+h_{\mu\nu}(x,y),
\end{equation}
where $|h_{\mu\nu}\ll1|$.
To first order in $h$, we obtain the Ricci scalar of the metric (\ref{flat GW}) by making use of the formula (\ref{Ricci scalar},\ref{geodesic spray})
\begin{eqnarray}
F^2Ric&=&\frac{1}{2}\eta^{\mu\nu}y^\alpha y^\beta\left(2\frac{\partial^2h_{\alpha\nu}}{\partial x^\mu \partial x^\beta}-\frac{\partial^2h_{\alpha\beta}}{\partial x^\mu \partial x^\nu}-\frac{\partial^2h_{\mu\nu}}{\partial x^\alpha \partial x^\beta}\right)\nonumber\\
\label{flat GW Ric}
&&-\frac{1}{4}y^\theta y^\alpha y^\beta \frac{\partial}{\partial y^\mu}\left(2\frac{\partial^2(\eta^{\mu\nu}h_{\alpha\nu})}{\partial x^\theta \partial x^\beta}-\frac{\partial^2(\eta^{\mu\nu}h_{\alpha\beta})}{\partial x^\nu \partial x^\theta}\right).
\end{eqnarray}
We consider the gravitational wave coming from infinity for simplicity, which means that gravitational source that produces the gravitational wave can be neglected. The discussion about the vacuum field equation in Finsler spacetime requires that such gravitational wave should satisfy $Ric=0$.  It is rather complicated to solve the equation $Ric=0$ for gravitational wave in Finsler spacetime. And we are only interested in Finslerian plane wave solution of equation $Ric=0$ in physics.

In order to get the plane wave solution of gravitational wave, we suggest three constraints on gravitational wave. The first one is $h^\mu_\nu=\eta^{\mu\alpha}h_{\alpha\nu}=h^\mu_\nu(x)$. The first constraint requires that $h^\mu_\nu$ is only a function of $x$. It means that we choose a special tensor perturbations for flat Finsler spacetime. And such special perturbation will reduce to standard tensor perturbations in general relativity if the metric of flat Finsler spacetime $\eta_{\mu\nu}(y)$ returns to Minkowski metric. The second one is the gauge condition
\begin{equation}\label{gauge}
\frac{2\partial h^\mu_\nu}{\partial x^\mu}-\frac{\partial h^\mu_\mu}{\partial x^\nu}=0.
\end{equation}
Such gauge condition is same to the one in general relativity. And it can be satisfied in Finsler spacetime, since the Ricci scalar is invariant under coordinate transformation. The last constraint states that the direction of $y$ is parallel with $\frac{\partial}{\partial x}$.  The Finslerian length element $F$ is constructed on a tangent bundle \cite{Book by Bao}. Thus, the gravitational field equation should be constructed on the tangent bundle in principle. The last constraint implies that we have restricted the field equation on base manifold such that the fiber coordinate $y$ is parallel to $\frac{\partial}{\partial x}$. By making use of the three constraints, and noticing the relation $\frac{\partial\bar{g}^{ij}}{\partial y^i}y_j=0$, we find that the formula (\ref{flat GW Ric}) reduces to
\begin{equation}\label{flat GW Ric1}
F^2Ric=-\frac{1}{2}\eta^{\mu\nu}(k)\frac{\partial^2h_{\alpha\beta}}{\partial x^\mu \partial x^\nu}y^\alpha y^\beta.
\end{equation}
Since $\eta^{\mu\nu}$ is homogenous function of degree $0$ with respect to variable $y$ and $y$ is parallel with wave vector $k$ in momentum space, we have replaced the variable $y$ of $\eta^{\mu\nu}$ into $k$ in formula (\ref{flat GW Ric1}).
Plugging the formula (\ref{flat GW Ric1}) into the field equation $Ric=0$, we obtain the solution of field equation
\begin{equation}\label{plane wave}
h^\mu_\nu=e^\mu_\nu\exp(i\eta_{\alpha\beta}(k)k^\alpha x^\beta)+h.c.~,
\end{equation}
where $e^\mu_\nu$ denotes the polarization tensor of gravitational wave and the wave vector $k$ satisfies
\begin{equation}\label{null k}
\eta_{\mu\nu}(k)k^\mu k^\nu=0.
\end{equation}
The equation (\ref{null k}) represents that the velocity of gravitational wave depends on wave vector $k$. It means that the Lorentz symmetry is violated that is a feature of Finsler spacetime \cite{Gibbons,Kostelecky}. The plane wave solution (\ref{plane wave}) satisfies the gauge condition (\ref{gauge}) if
\begin{equation}\label{gauge k}
2k_\mu e^\mu_\nu=k_\nu e^\mu_\mu,
\end{equation}
where $k_\mu=\eta_{\mu\nu}k^\nu$. The four relations in (\ref{gauge}) imply that the polarization tensor $e^\mu_\nu$ have six independent components. Following the approach in general relativity \cite{Weinberg}, one could find that only two components of polarization tensor are physical.

\section{Gravitational wave in Finslerian inflation}\label{sec:FRW GW}
In Ref.\cite{Finsler scalar modes}, we propose a background Finsler spacetime to describe the anisotropic inflation. It is of the form
\begin{equation}\label{FRW like}
F^2_0=y^ty^t-a^2(t)F_{Ra}^2,
\end{equation}
where $F_{Ra}$ is a Randers space \cite{Randers}
\begin{equation}
F_{Ra}=\sqrt{\delta_{ij}y^iy^j}+\delta_{ij}b^iy^j.
\end{equation}
Here, we require that the vector $b^i$ in $F_{Ra}$ is of the form $b^i=\{0,0,b\}$ and $b$ is a constant. The spatial part of Finsler spacetime (\ref{FRW like}), i.e. $F_{Ra}$, preserves three translation symmetry and one rotational symmetry \cite{Finsler PF,Finsler BH,Finsler scalar modes}. It means that $F_{Ra}$ is only rotational invariant on the plane that is perpendicular to vector $b^i$, and other rotational symmetry of Euclidean space are broken. In this paper, we focus on investigating the tensor perturbations of the background Finsler spacetime (\ref{FRW like}). The perturbed Finsler structure is of the form
\begin{equation}\label{FRW GW}
F^2=y^ty^t-a^2(t)(F_{Ra}^2+h_{ij}y^i y^j).
\end{equation}
Here, we require the perturbed metric $h_{ij}$ satisfies the first and third constraint as discussed in section \ref{sec:flat GW}. And the second constraint, i.e. the gauge condition, is changed into
\begin{equation}
h^i_i=h^i_{j,i}=0,
\end{equation}
where the comma denotes the derivative with respect to spatial coordinate $\vec{x}$. By making use of the three constraints, we obtain the Ricci scalar of the perturbed Finsler spacetime (\ref{FRW GW}) in momentum space
\begin{eqnarray}
F^2Ric&=&(a\ddot{a}+2\dot{a}^2)(F^2_{Ra}+h_{ij} y^i y^j)-\frac{3\ddot{a}}{a}y^ty^t\nonumber\\
\label{Ricci scalar1}
&&+\frac{a^2}{2}\left(\ddot{h}_{ij}+\frac{3\dot{a}}{a}\dot{h}_{ij}+\frac{1}{a^2}\eta^{mn}k_m k_nh_{ij}\right)y^i y^j,
\end{eqnarray}
where the dot denotes the derivative with respect to time and $\eta^{mn}$ denotes the Finslerian metric of Randers space $F_{Ra}$.

In reference \cite{Finsler BH,Finslerian dipole}, we have proved that the gravitational field equation in Finsler spacetime
\begin{equation}\label{field equation}
G^\mu_\nu=8\pi G T^\mu_\nu
\end{equation}
is valid for the modified FRW spacetime (\ref{FRW like}) and Finslerian Schwarzschild spacetime \cite{Finsler BH}.
Here the modified Einstein tensor in Finsler spacetime is defined as
\begin{equation}\label{Einstein tensor}
G^\mu_\nu\equiv Ric^\mu_\nu-\frac{1}{2}\delta^\mu_\nu S,
\end{equation}
and $T^\mu_\nu$ is the energy-momentum tensor.
Here the Ricci tensor is defined as \cite{Akbar}
\begin{equation}\label{Ricci tensor}
Ric_{\mu\nu}=\frac{\partial^2\left(\frac{1}{2}F^2 Ric\right)}{\partial y^\mu\partial y^\nu},
\end{equation}
and the scalar curvature in Finsler spacetime is given as $S=g^{\mu\nu}Ric_{\mu\nu}$. Plugging the equation for Ricci scalar (\ref{Ricci scalar1}) into the gravitational field equation (\ref{field equation}), we obtain the perturbed equation for $h^i_j$
\begin{equation}\label{GW field equation}
-\frac{1}{2}\left(\ddot{h}^i_j+\frac{3\dot{a}}{a}\dot{h}^i_j+\frac{k^2_e}{a^2}h^i_j\right)=8\pi G\delta T^i_j,
\end{equation}
where $\delta T^i_j=0$ denotes the first order part of energy-momentum tensor and the effective wavenumber $k_e$ is given by
\begin{equation}
k_e^2=\eta^{ij}k_ik_j=k^2(1+b\hat{k}\cdot\hat{n}_z)^2.
\end{equation}
Here, $\hat{k}$ denotes the propagational direction of the gravitational wave.
The perturbed equation (\ref{GW field equation}) is same to the one in the standard inflation model except for replacing wavenumber $k$ with effective wavenumber $k_e$. $k_e$ depends not only on the magnitude of $k$ but also the preferred direction $\hat{n}_z$ that induces rotational symmetry breaking. Then, following the standard quantization process in the inflation model \cite{Mukhanov book}, we can obtain the primordial power spectrum of tensor perturbations from the solution of the equation (\ref{GW field equation}). It is of the form
\begin{equation}\label{spectrum h}
\mathcal{P}^{\pm2\pm2}(\vec{k})=\mathcal{P}^{\pm2\pm2}_{iso}(k)\left(\frac{k}{k_e}\right)^3\simeq\mathcal{P}^{\pm2\pm2}_{iso}(k)(1-3b\hat{k}\cdot\hat{n}_z),
\end{equation}
where $\mathcal{P}^{\pm2\pm2}_{iso}$ is an isotropic power spectrum for tensor perturbations $h$ which depends only on the magnitude of wavenumber $k$.
The term $3b\hat{k}\cdot\hat{n}_z$ in the primordial power spectrum $\mathcal{P}^{\pm2\pm2}$ represents the effect of rotational symmetry breaking.

\begin{figure}
\includegraphics[width=8.5 cm]{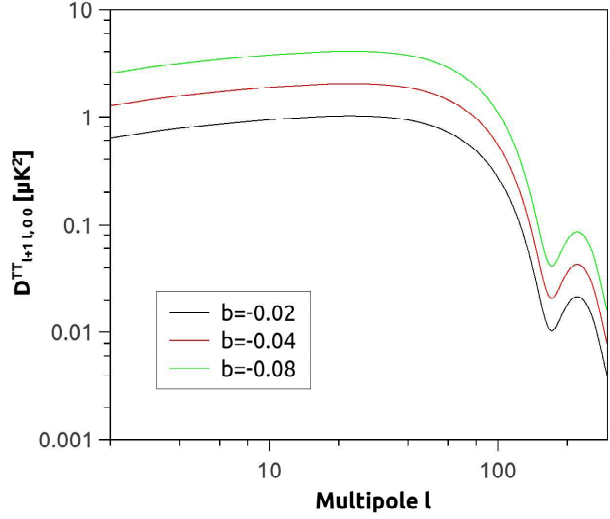}
\caption{The anisotropic part of $TT$ correlation coefficients with $m=0$. The black, red and green curves correspond to different Finslerian parameter, i.e. $b=-0.02,-0.04,-0.08$, respectively. The following figures are given by the same convention.}
\label{TT m=0}
\end{figure}
\begin{figure}
\includegraphics[width=8.5 cm]{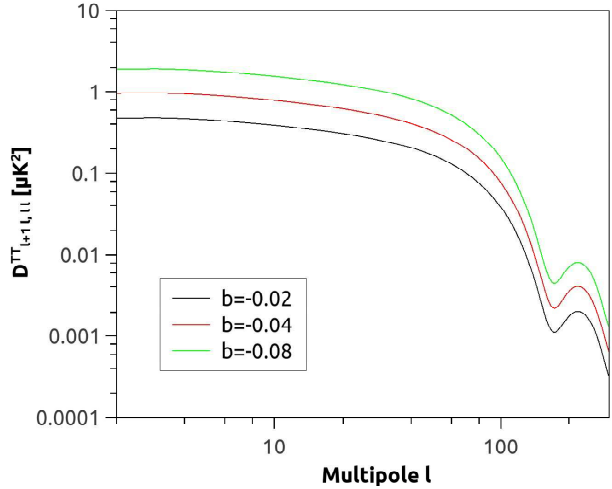}
\caption{The anisotropic part of $TT$ correlation coefficients with $m=l$. It demonstrates that the anisotropic part of correlation coefficients $C_{ll'}$ depend on $m$.}
\label{TT m=l}
\end{figure}
\begin{figure}
\includegraphics[width=8.5 cm]{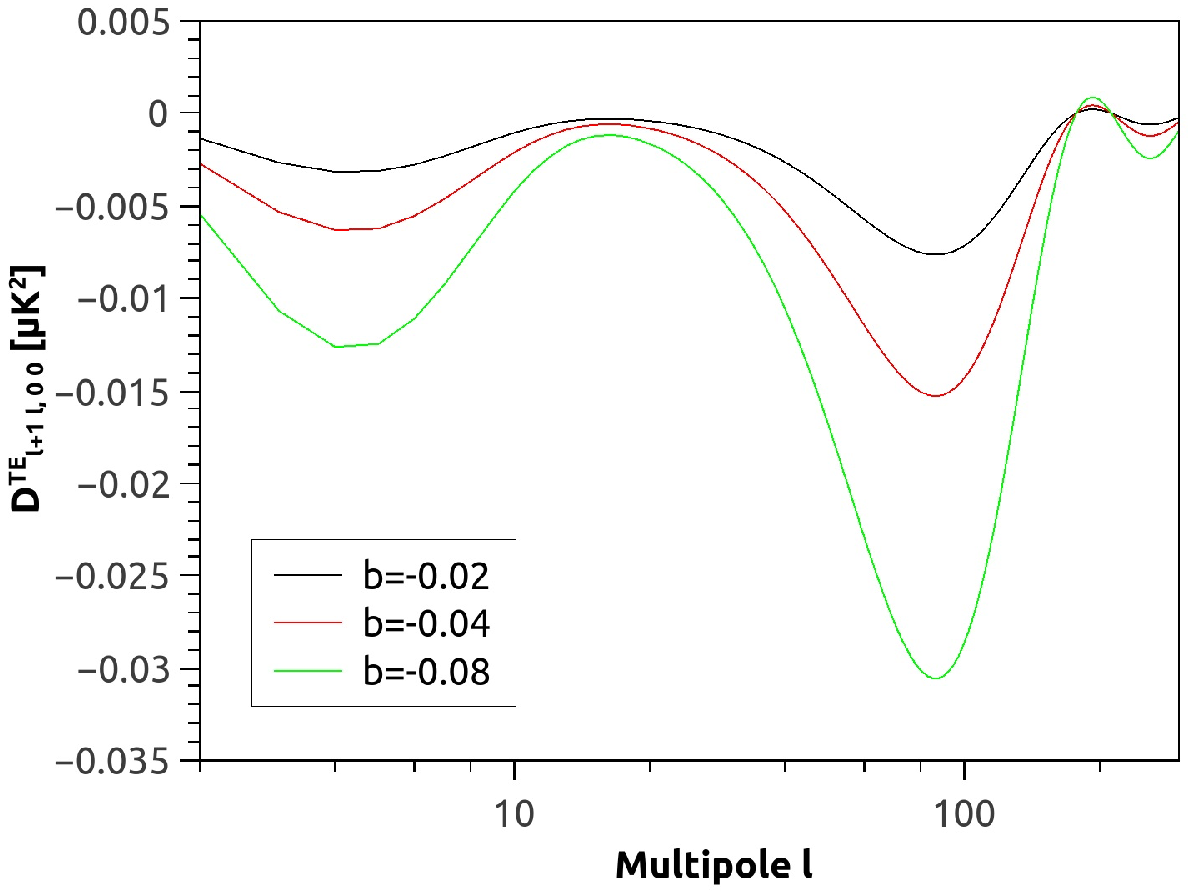}
\caption{The anisotropic part of $TE$ correlation coefficients with $m=0$.}
\label{TE m=0}
\end{figure}
\begin{figure}
\includegraphics[width=8.5 cm]{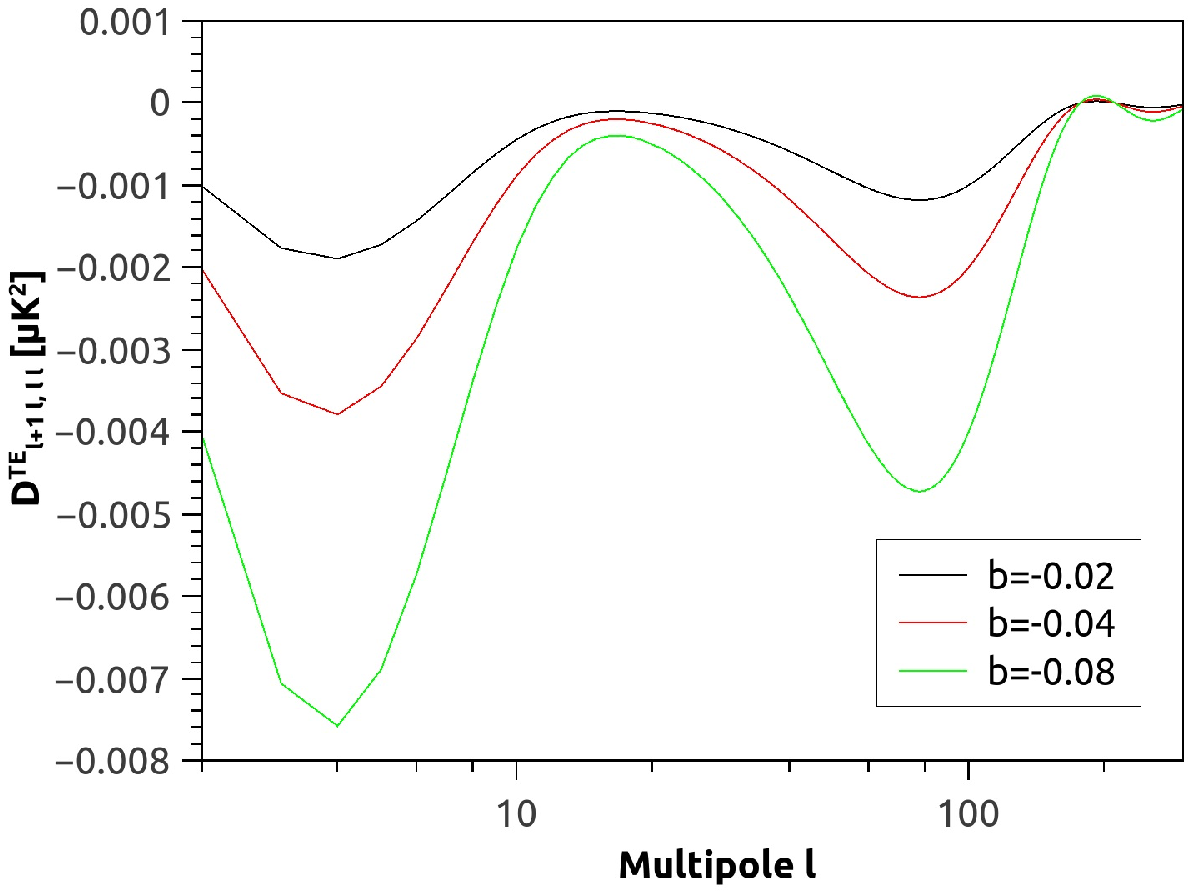}
\caption{The anisotropic part of $TE$ correlation coefficients with $m=l$.}
\label{TE m=l}
\end{figure}
\begin{figure}
\includegraphics[width=8.5 cm]{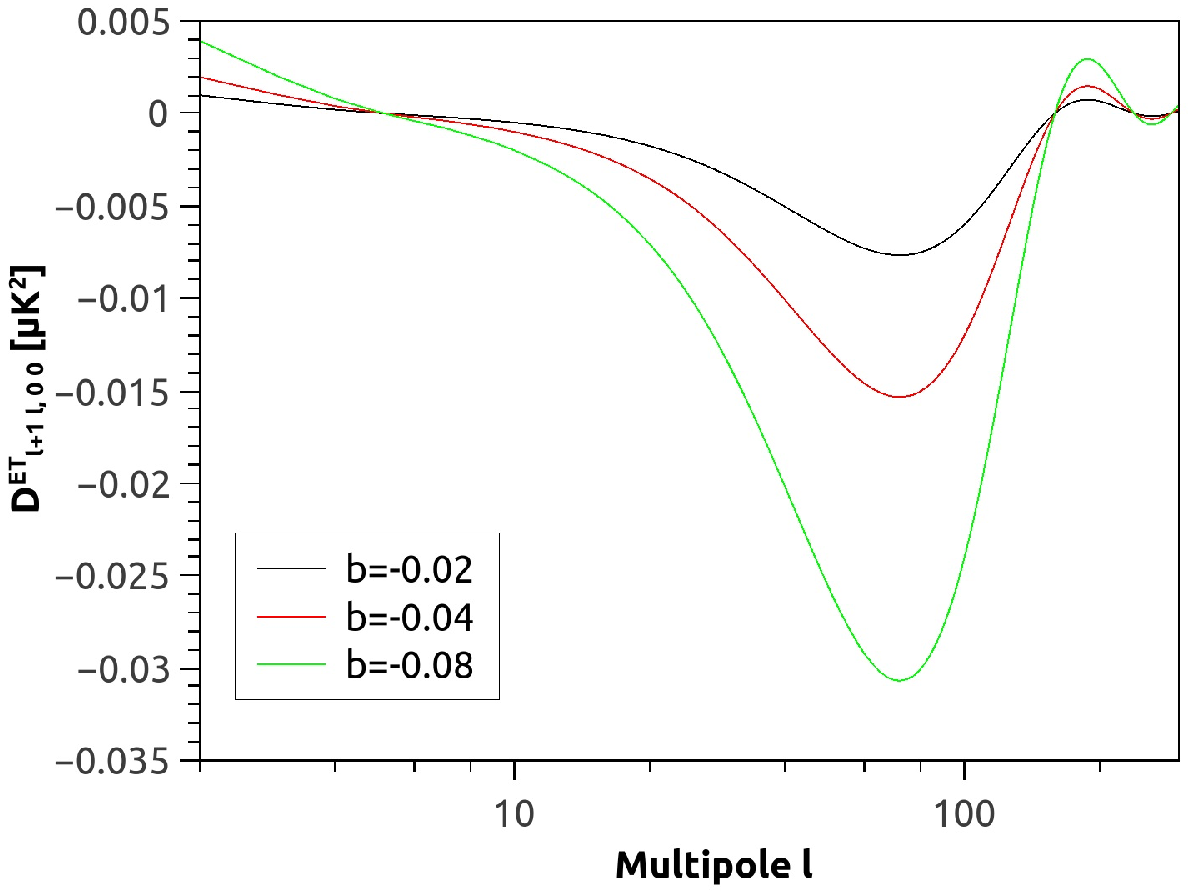}
\caption{The anisotropic part of $ET$ correlation coefficients with $m=0$.}
\label{ET m=0}
\end{figure}
\begin{figure}
\includegraphics[width=8.5 cm]{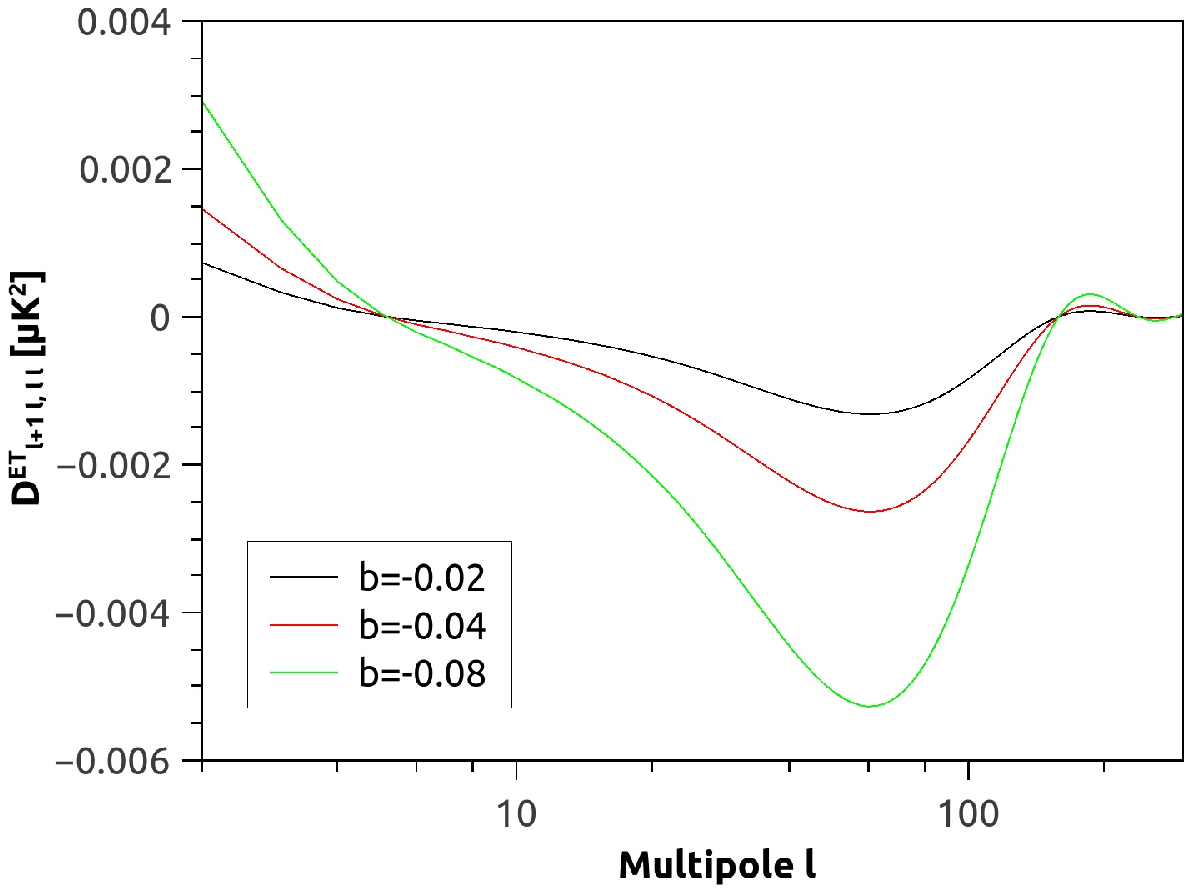}
\caption{The anisotropic part of $ET$ correlation coefficients with $m=l$.}
\label{ET m=l}
\end{figure}
\begin{figure}
\includegraphics[width=8.5 cm]{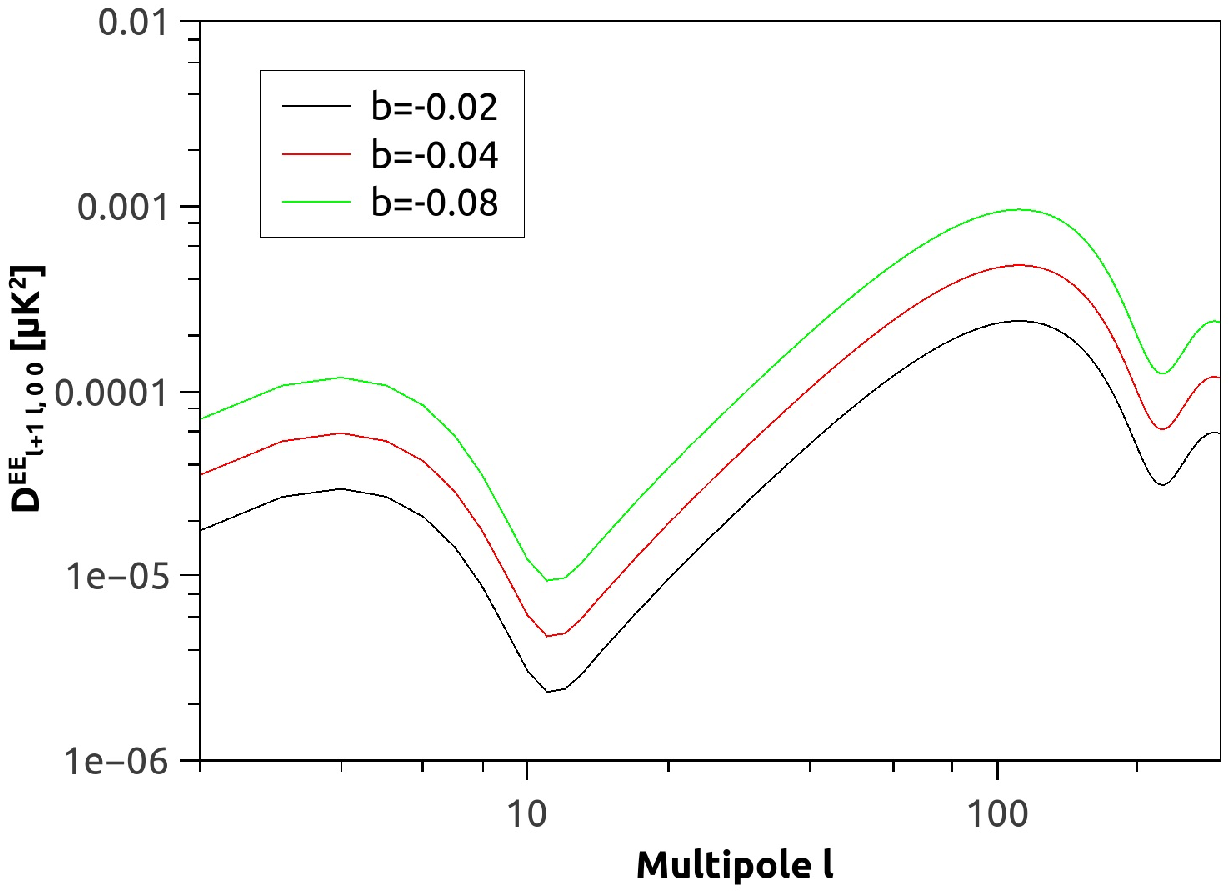}
\caption{The anisotropic part of $EE$ correlation coefficients with $m=0$.}
\label{EE m=0}
\end{figure}
\begin{figure}
\includegraphics[width=8.5 cm]{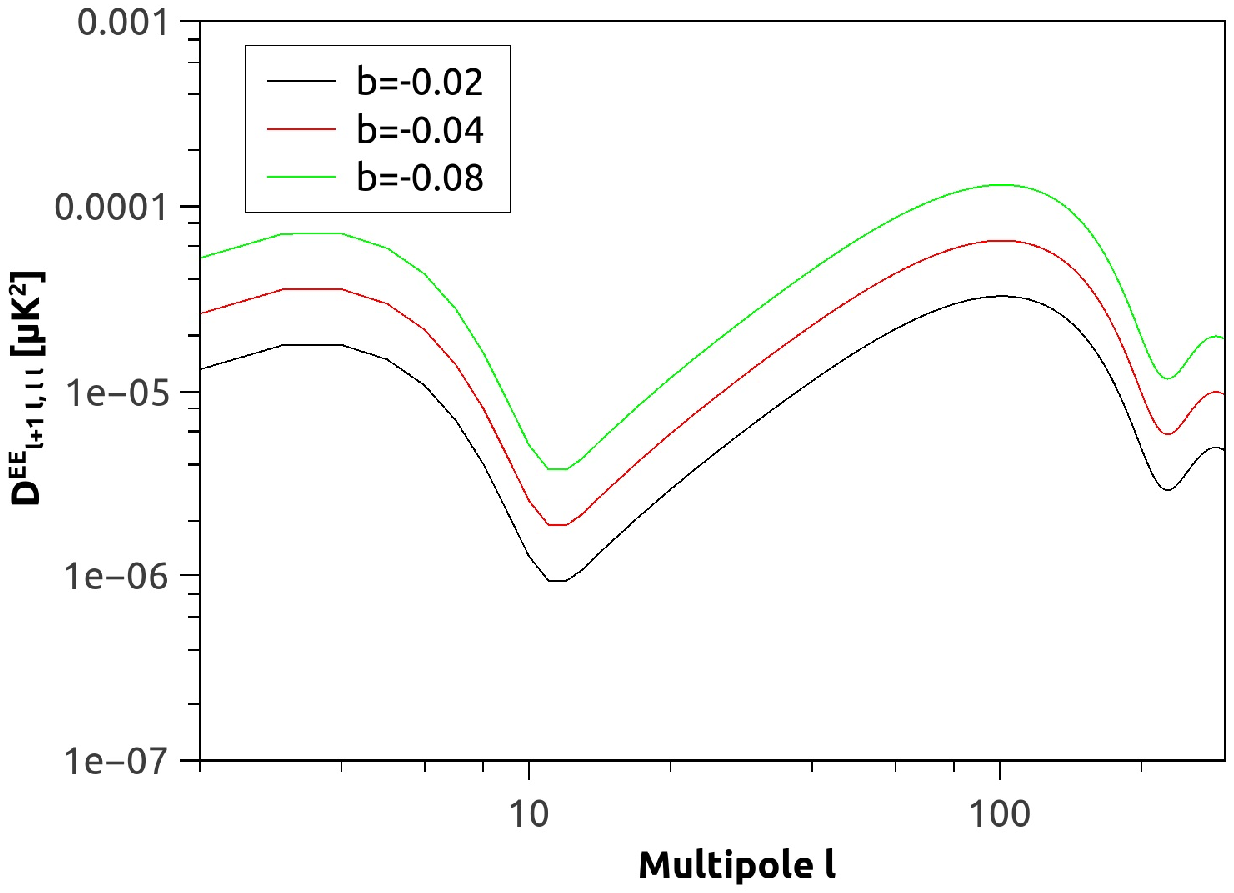}
\caption{The anisotropic part of $EE$ correlation coefficients with $m=l$.}
\label{EE m=l}
\end{figure}
\begin{figure}
\includegraphics[width=8.5 cm]{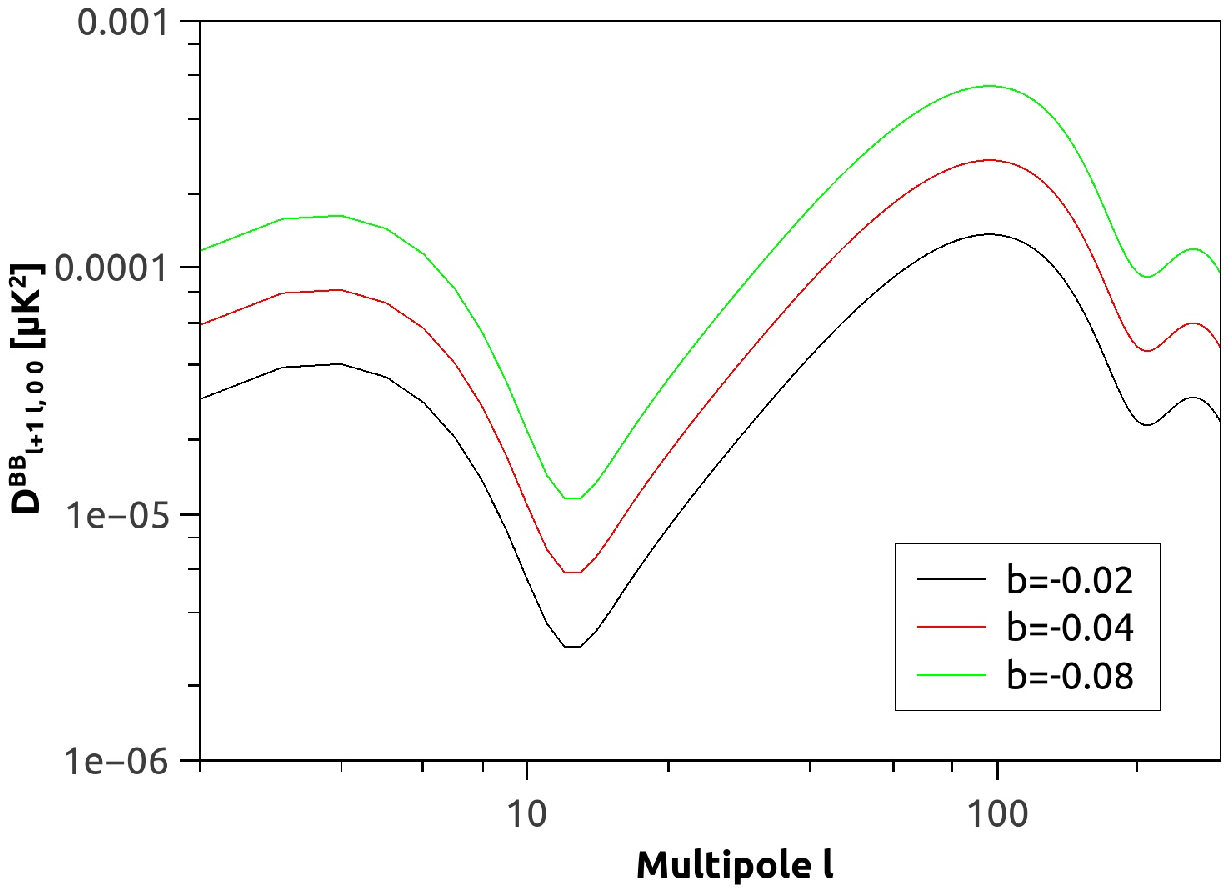}
\caption{The anisotropic part of $BB$ correlation coefficients with $m=0$.}
\label{BB m=0}
\end{figure}
\begin{figure}
\includegraphics[width=8.5 cm]{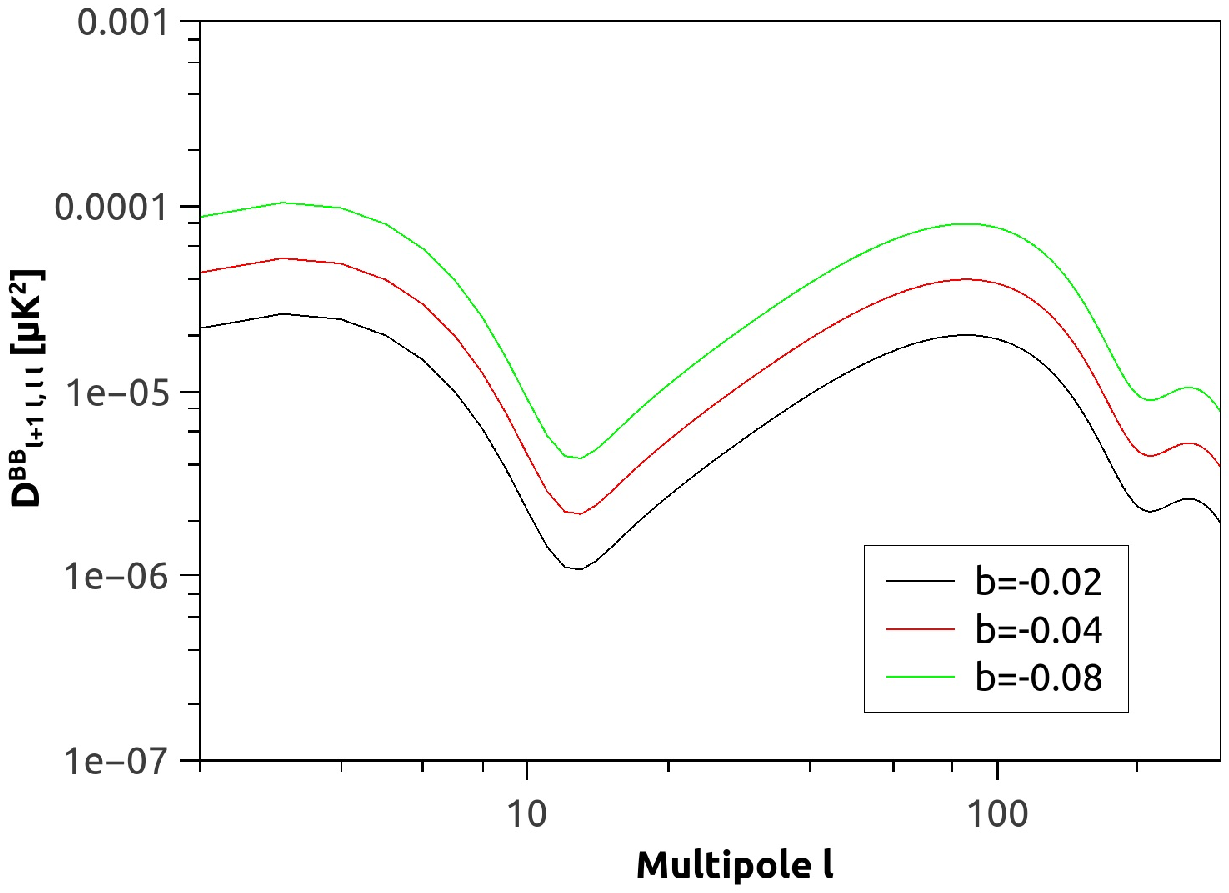}
\caption{The anisotropic part of $BB$ correlation coefficients with $m=l$.}
\label{BB m=l}
\end{figure}
\begin{figure}
\includegraphics[width=8.5 cm]{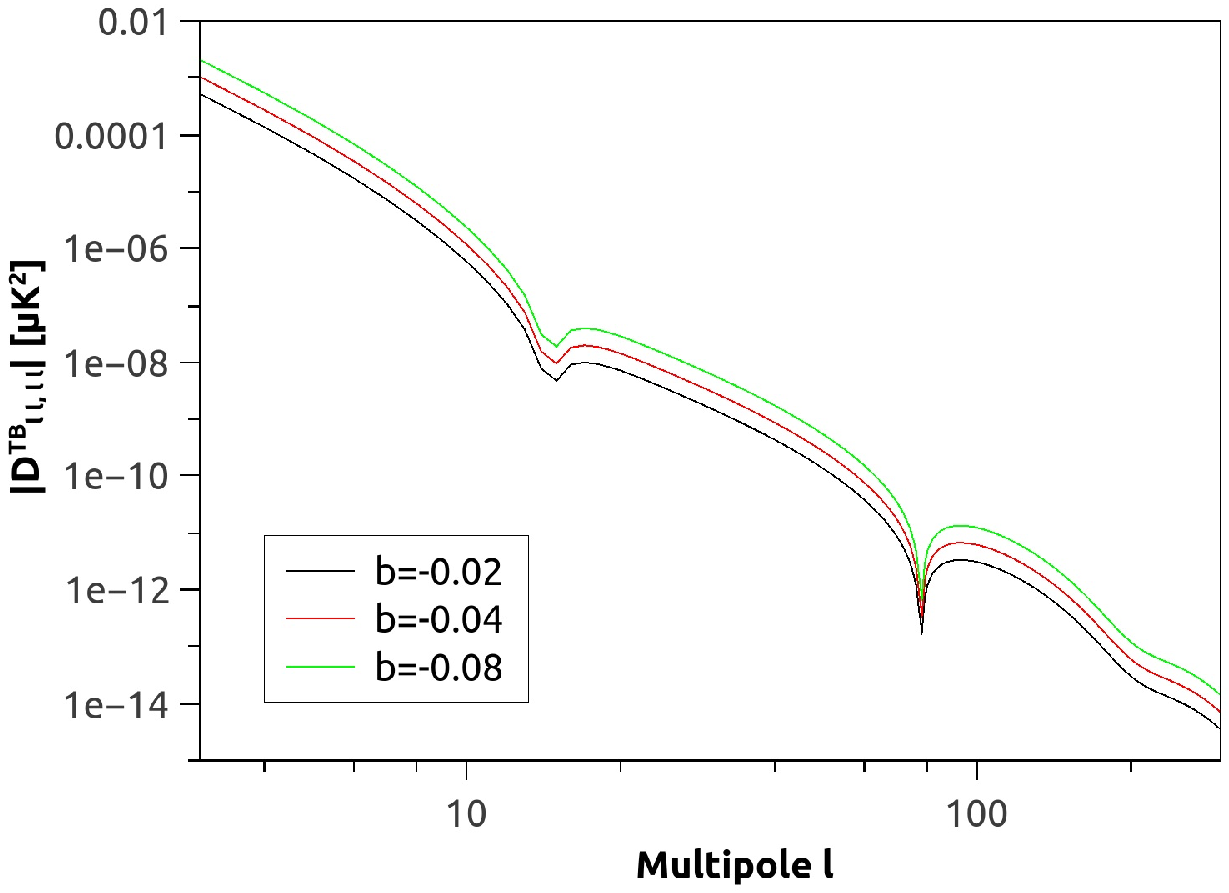}
\caption{The anisotropic part of $TB$ correlation coefficients with $m=l$.}
\label{TB m=l}
\end{figure}
\begin{figure}
\includegraphics[width=8.5 cm]{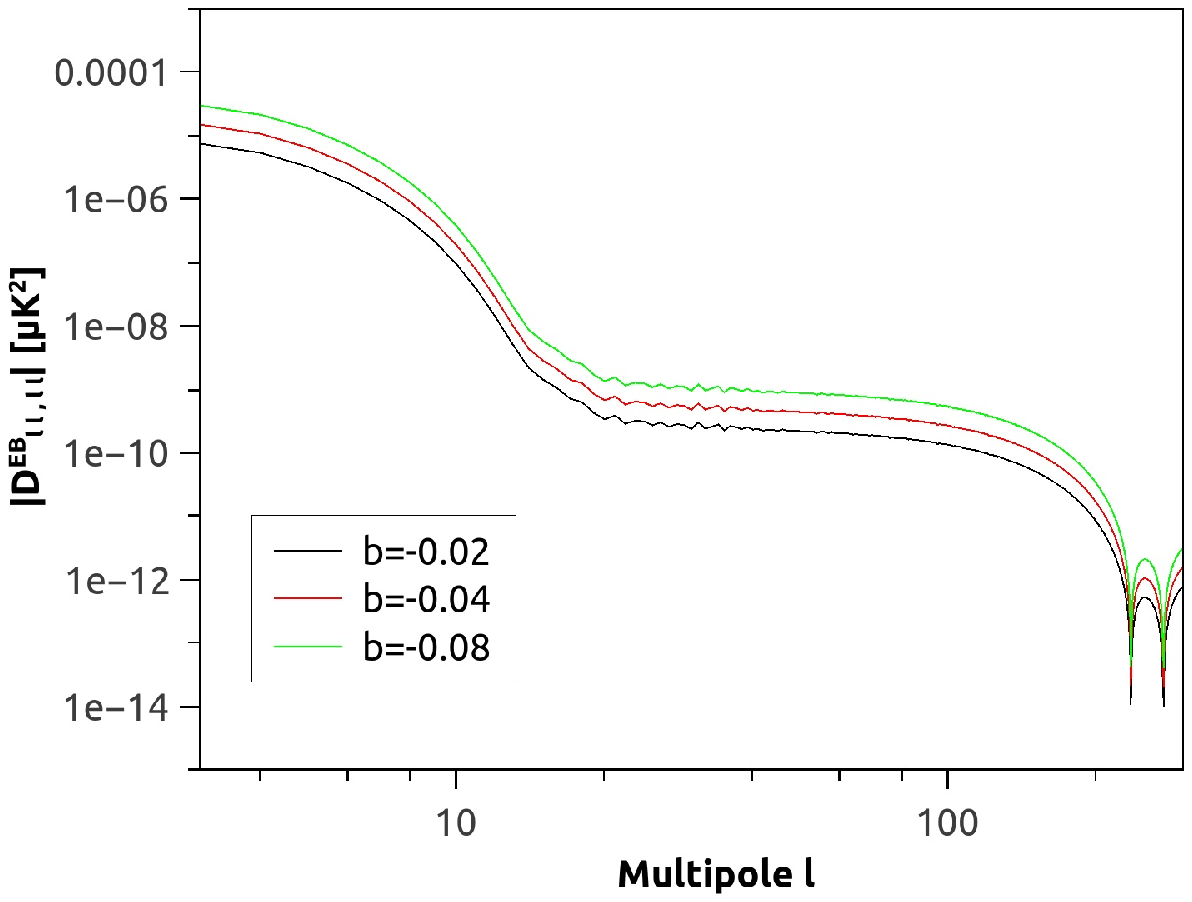}
\caption{The anisotropic part of $EB$ correlation coefficients with $m=l$.}
\label{EB m=l}
\end{figure}

\section{The anisotropic effects on angular power spectra}\label{sec:numerical results}
The anisotropic term in formula (\ref{spectrum h}) could give off-diagonal angular correlation for the CMB temperature fluctuation, E-mode and B-mode polarization of CMB, and it also contributes to the $TB$ and $EB$ spectra that should vanish in standard inflation model. The general angular correlation coefficients for tensor perturbations that describe the anisotropic effect are given by $C_{XX',ll',mm'}$ \cite{off-diagonal,off-diagonal1}, where $X=T,E,B$ denotes CMB temperature fluctuation, E-mode and B-mode polarizations, respectively. In our anisotropic model, by making use of the formula (\ref{spectrum h}), we obtain the CMB correlation coefficients for tensor perturbations as follows
\begin{equation}\label{angular correlation}
C^T_{XX',ll',mm'}=\int\frac{d\ln k}{(2\pi)^3}\Delta_{X,l2}(k)\Delta^\ast_{X',l'2}(k)P^{\pm2\pm2}_{ll'mm'},
\end{equation}
where
\begin{eqnarray}\label{angular correlation1}
P^{\pm2\pm2}_{ll'mm'}&=&\int d\Omega\mathcal{P}^{\pm2\pm2}(\vec{k})\left(_{-2}Y^\ast_{lm}~_{-2}Y_{l'm'}\pm~_{+2}Y^\ast_{lm}~_{+2}Y_{l'm'}\right)\nonumber\\
&=&\mathcal{P}^{\pm2\pm2}_{iso}\sqrt{\frac{2l'+1}{2l+1}}\left(\mathcal{C}^{l'm'}_{00lm}\left(\mathcal{C}^{l'2}_{00l2}\pm\mathcal{C}^{l'-2}_{00l-2}\right)-3b\mathcal{C}^{l'm'}_{10lm}\left(\mathcal{C}^{l'2}_{10l2}\pm\mathcal{C}^{l'-2}_{10l-2}\right)\right),
\end{eqnarray}
$\Delta_{X,l2}(k)$ denote the transfer functions for tensor modes and $\mathcal{C}^{l'm'}_{LMlm}$ are the Clebsch-Gordan coefficients. Here, $_sY_{lm}$ in the formula (\ref{angular correlation1}) are the spin-$s$ spherical harmonic function \cite{spin weighted harmonic}. In the formula (\ref{angular correlation1}), the `$+$' of `$\pm$' in bracket corresponds to $TT,TE,EE,BB$ correlations and the `$-$' of `$\pm$' corresponds to $TB,EB$ correlations. By making use of the symmetry of Clebsch-Gordan coefficients $\mathcal{C}^{l'm'}_{LMlm}=(-1)^{L+l-l'}\mathcal{C}^{l'-m'}_{L-Ml-m}$, one can find from the formula (\ref{angular correlation}) that $TT,TE,EE,BB$ correlations have non-zero value for $l'=l,l+1$ and $TB,EB$ correlations have non-zero value for $l'=l$. The anisotropic term $3b\hat{k}\cdot\hat{n}_z$ in the primordial power spectrum $\mathcal{P}^{\pm2\pm2}$ that described the deviation from statistical isotropy violates the parity symmetry. Thus, it contributes to $TT,TE,EE,BB$ correlations if $l'=l+1$ and $TB,EB$ correlations if $l'=l$.

Here, by making use of the formula of angular correlation coefficients (\ref{angular correlation}) and the Planck 2015 data \cite{Planck XIII}, we plot numerical results for the anisotropic contribution to $C_{ll'}\equiv \mathcal{C}_{XX',ll',mm'}$. The anisotropic part of $C_{ll'}$ have three properties that differ from the isotropic part. The first one is that the $TE$ and $ET$ correlation coefficients are different. The second one shows that the anisotropic part of $C_{ll'}$ depends on $m$. The last one shows that $C^{TB}_{ll}$ and $C^{EB}_{ll}$ have non--vanishing value. These properties are obvious in the following Fig.~\ref{TT m=0}--\ref{EB m=l}. At present, the observations of CMB, such as the Planck data \cite{Planck XI}, do not give $TT,TE,EE,BB$ correlations for $l'=l+1$ and $TB,EB$ correlations for $l'=l$. Thus, to show the effect of the Finslerian modification for primordial tensor perturbations, we set Finslerian parameter to be $b=-0.02,-0.04,-0.08$ that have the same order with the magnitude of the CMB dipole modulation \cite{Planck XVI}. The anisotropic part of $TT,TE,ET,EE$ correlation coefficients for $m=0$ are shown in Fig.~\ref{TT m=0}, \ref{TE m=0}, \ref{ET m=0}, \ref{EE m=0}, \ref{BB m=0}, respectively. And the anisotropic contributions to $TT,TE,ET,EE$ correlation coefficients for $m=l$ are shown in Fig.~\ref{TT m=l}, \ref{TE m=l}, \ref{ET m=l}, \ref{EE m=l}, \ref{BB m=l}, respectively. The anisotropic contributions to $TB$ and $EB$ are shown in Fig.~\ref{TB m=l}, \ref{EB m=l}. Here, we have used the mean value of cosmological parameters \cite{Planck XIII} to give the above figures. And the tensor-to-scalar $r$ is set to be $r=0.1$ that is compatible with the current observations \cite{BKP}. The coefficients $D^{XX'}_{ll',mm'}$ in these figures are defined as $D^{XX'}_{ll',mm'}\equiv (2\pi)^{-1}\sqrt{l(l+1)l'(l'+1)}C_{XX',ll',mm'}$. Here the pivot scale is set to be $k_p=0.01~\rm{Mpc}^{-1}$ for tensor perturbations.

\section{Conclusions and Remarks}\label{sec:conclusion}
In this paper, we have investigated the gravitational wave in the flat Finsler spacetime. To get the plane wave solution of gravitational wave, three constraints are involved. In the modified FRW spacetime (\ref{FRW GW}) with tensor perturbations, we derived the perturbed gravitational field equation for tensor perturbations (\ref{GW field equation}) by making use of the three constraints. From the solution of the perturbed gravitational field equation, we obtain the primordial power spectrum of tensor perturbations (\ref{spectrum h}). The term $3b\hat{k}\cdot\hat{n}_z$ in the primordial power spectrum (\ref{spectrum h}) that violates the rotational symmetry and parity symmetry describes the statistical anisotropy of CMB temperature fluctuation, E-mode and B-mode polarizations. We have used the primordial power spectrum (\ref{spectrum h}) to derive the angular correlation coefficients $C_{XX',ll',mm'}$. The parity violation feature requires that the anisotropic effect appears in $TT,TE,EE,BB$ correlations for $l'=l+1$ and $TB,EB$ correlations for $l'=l$. The numerical results for the anisotropic parts of the correlation coefficients show that they depend on $m$, and $TE$ and $ET$ correlation coefficients are different.

\vspace{1cm}
\begin{acknowledgments}
X.Li has been supported by the National Natural Science Fund of China (NSFC) (Grant NO. 11305181) and the Open Project Program of State Key Laboratory of Theoretical Physics, Institute of Theoretical Physics, Chinese Academy of Sciences, China (No. Y5KF181CJ1). S.Wang has been supported by grants from NSFC (Grant NO. 11322545 and 11335012).
\end{acknowledgments}

\end{document}